\documentclass[twoside]{article}
\usepackage{fleqn,espcrc2}

\usepackage{graphicx}

\title{Impurity spin dynamics in 2D antiferromagnets and superconductors}

\author{ Matthias Vojta, Chiranjeeb Buragohain and Subir Sachdev
\address{Department of Physics, Yale University, P.O.
         Box 208120, New Haven, CT 06520-8120, USA}%
        \thanks{Research supported by US NSF Grant No
                DMR 96--23181 and by the DFG (VO 794/1-1)}
        }

\begin{document}

\begin{abstract}

We discuss the universal theory of localized impurities in the paramagnetic state
of 2D antiferromagnets
where the spin gap is assumed to be significantly smaller than a
typical exchange energy.
We study the impurity spin susceptibility
near the host quantum transition from a gapped paramagnet to a
N\'{e}el state,
and we compute the impurity-induced damping of the spin-1
mode of the gapped antiferromagnet.
Under suitable conditions our results apply also to d-wave superconductors.
\end{abstract}

\maketitle


Doped antiferromagnets (AF) have been the subject of intense studies in the context of the
cuprate high-temperature superconductors and other layered
transition metal compounds.
We present a quantum theory of a particular class of doped AF
where it is possible to neglect the coupling between
the spin and charge degrees of freedom and consider a theory of
the spin excitations alone.
Such a theory will apply to
(i) quasi-2D `spin gap' insulators
like ${\rm Sr Cu}_2 {\rm O}_3$ or ${\rm Na V}_2 {\rm O}_5$
in which a small fraction of the magnetic ions
(Cu or V) are replaced by non-magnetic ions like Zn or
Li and to
(ii) high-temperature superconductors like ${\rm Y} {\rm Ba}_2 {\rm
Cu}_3 {\rm O}_7$ in which a small fraction of Cu has been replaced
by non-magnetic Zn or Li.
In the first case the spin gap $\Delta$ is significantly smaller than the charge
gap justifying a theory of the spin excitations alone.
In the second situation the effect of the fermionic quasiparticles in
the superconducting state can be shown \cite{long} to be weak
due to the linearly vanishing density of states of the Fermi level.

The effect of a (magnetic or non-magnetic) impurity can be probed by
measuring the uniform spin susceptibility, which takes the form
$\chi = (g \mu_B)^2 (A \chi_b + \chi_{{\rm imp}})$ where $A$ is
the total area of the AF,
$\chi_b$ is the bulk response per unit area, and $\chi_{{\rm imp}}$ is the additional
impurity contribution.
In the paramagnetic ground state of the host
each impurity induces a distortion of the host spin arrangement
with a net magnetic moment $S$ associated with
the impurity.
The distortion is {\em confined} to the vicinity of the impurity
which implies that the impurity susceptibility follows
\begin{equation}
\chi_{{\rm imp}} = \frac{S(S+1)}{3 k_B T} ~~~~ \mbox{as}~~T\to 0.
\label{chi1}
\end{equation}
For a non-magnetic impurity in a spin-1/2 system we have $S=1/2$;
for a general impurity eq. (\ref{chi1}) can be used as definition
of $S$.

The basis of our investigations is a boundary quantum field theory
which describes a bulk AF together with arbitrary
localized deformations.
We focus on the vicinity of a quantum transition from a paramagnet to a magnetically
ordered N\'{e}el state:
Then the spin gap in the paramagnetic state is small
compared to a typical nearest-neighbor exchange, $\Delta \ll J$, which
is the situation realized in many compounds.
The field theory has been discussed in Ref. \cite{long};
it consists of $d+1$-dimensional $\phi^4$ theory for the bulk ordering
transition and a coupling to a local quantum impurity spin.
The renormalization-group (RG) analysis shows that both the bulk and the boundary couplings
are marginal for $d=3$ and flow to fixed-point values for $d<3$.
This implies that the coupling between the bulk and impurity excitations
becomes {\em universal},
and the spin dynamics in the vicinity of the impurity is completely determined
by bulk parameters, the gap $\Delta$ and the velocity of spin
excitations $c$.
Based on the RG results one obtains a number of universal properties in an
expansion in $\epsilon=3-d$; we mention here the behavior of the uniform susceptibility
at the bulk critical point.
The system shows the Curie response of an {\em irrational} spin
as $T\to 0$, $\chi_{{\rm imp}} = {\cal C}_1 / (k_B T)$,
where ${\cal C}_1$ is a {\em universal} number independent of
microscopic details.
The $\epsilon$ expansion result for ${\cal C}_1$ is
\begin{equation}
{\cal C}_1 = \frac{S(S+1)}{3} \left[ 1 + \left(\frac{33 \epsilon}{40}\right)^{1/2}
\!\! -\frac{7 \epsilon}{4} + ...   \right].
\end{equation}

\newcommand{\bb}  {{f}}         
\newcommand{\ttd} {{t}}         
\newcommand{\KK}  {{K}}         
\newcommand{\JJ}  {{J}}         
\newcommand{\KKK} {{\tilde K}}  
\newcommand{\OM}  {{\epsilon}}  
\newcommand{\MM}  {m}           
\newcommand{\DD}  {\Delta}      
\newcommand{\coup} {s}          

More detailed dynamic information can be obtained by a self-consistent
diagrammatic method.
The paramagnetic phase of the bulk is assumed to be dimerized,
its spin-1 excitations can be described using triplet bosons $\ttd_{\bf k \alpha}$.
The impurity is represented by an additional spin $S_\alpha$ at site 0,
\begin{eqnarray}
H =
\sum_{{\bf k},\alpha} \OM_{\bf k} \ttd_{\bf k \alpha}^\dagger \ttd_{\bf k \alpha}
\,+\,
\frac{\KK}{\sqrt{N_s}}
\sum_{\bf k\alpha} S_\alpha
{ \ttd_{\bf k\alpha}^\dagger + \ttd_{\bf k\alpha}
\over
\sqrt{\OM_{\bf k} / \JJ } }
\end{eqnarray}
where $J$ is the host exchange constant, $\OM_{\bf k}$ the energy of the spin-1
mode in the bulk, $K$ the coupling constant to the impurity spin, and $N_s$ the number
of lattice sites.
The impurity spin is represented by auxiliary fermions $\bb$,
the impurity dynamics is contained in the fermion self-energy which arises
from the scattering off the $\ttd$ bosons.
We employ a self-consistent non-crossing approximation (NCA) to calculate
this self-energy; this approach follows from a saddle-point principle
after generalizing the spin symmetry to SU($N$) and taking
the limit $N\to\infty$.
The NCA equations can be solved in the scaling limit;
the value of the coupling $K$ drops out of all results for physical
observables provided that $\Delta\ll J$ -- we obtain the same universal
behavior as predicted by the RG.
In fact, the results for susceptibility and impurity spin correlations
agree with the one-loop RG result \cite{long}.

\begin{figure}
\centerline{\includegraphics[width=2.9in]{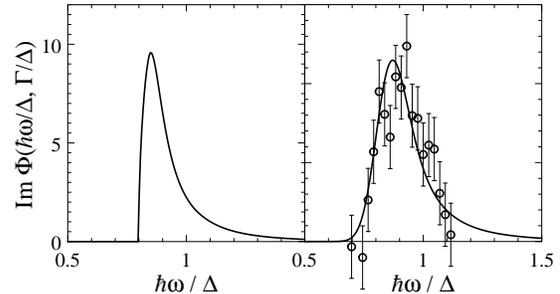}}
\vspace{-0.3in}
\caption{Left: Universal lineshape ${\rm Im} \Phi$
for $\Gamma/\Delta = 0.125$.
Right: The same, but convoluted with a Gaussian corresponding to the experimental
resolution of \protect\cite{keimer}, together with the data points of
Ref. \protect\cite{keimer}.
We have used $\Delta = 43$ meV -- this small shift may be attributed to
perturbations being irrelevant in the RG sense.
}
\vspace{-0.1in}
\label{fig2}
\end{figure}

The diagrammatic approach can be easily applied to a system with a finite
density of impurities $n_{\rm imp}$.
The important observation is that the impact of the impurities is
determined by a single energy scale
$\Gamma \equiv n_{\rm imp} (\hbar c)^d / \Delta^{d-1}$.
The AF in the absence of impurities shows a pole in the
dynamic susceptibility $\chi_{{\bf Q}} ( \omega )$ at the
AF wavevector $\bf Q$.
Our main concern is the fate of this collective peak upon the introduction
of impurities.
Scaling arguments predict that the susceptibility takes the form
\begin{equation}
\chi_{{\bf Q}} ( \omega ) = \frac{\mathcal{A}}{\Delta^2}
\Phi \left( \frac{\hbar \omega}{\Delta}, \frac{\Gamma}{\Delta}
\right)
\label{polescale} ~~~(T=0)\,,
\end{equation}
where $\Phi$ is a universal function, and $\cal A$ denotes the
quasiparticle weight.
In the absence of impurities we have
$\Phi ( \overline{\omega}, 0) = 1/(1 - (\overline{\omega} + i 0^{+})^2)$.
The self-energy of the spin-1 bosons caused by the scattering at
randomly distributed impurities is calculated using
a self-consistent Born approximation.
The equations for the Green's functions can be entirely
written in terms of scaling functions with arguments
$\hbar\omega/\Delta$ and $\Gamma/\Delta$, consistent with
the scaling prediction (\ref{polescale}).
A numerical result for $\Phi$ is shown in Fig.~\ref{fig2}.
The quasiparticle pole is broadened to an asymmetric line,
with a tail at high frequencies.
Our theory can be applied to a recent experiment \cite{keimer}
where the inpurity-induced broadening of the spin-1 `resonance peak'
at energy $\Delta = 40$ meV in ${\rm Y} {\rm Ba}_2 {\rm Cu}_3 {\rm O}_7$
has been observed.
This experiment has $n_{\rm imp}=0.005$, $\Gamma = 5$ meV, $\Gamma/\Delta = 0.125$.
The half-width of the line is approximately $\Gamma$, and this is
in excellent accord with the measured linewidth of 4.25 meV, see
Fig.~\ref{fig2}.
More tests of the predictions of our theory should be possible
in the future.

\end{document}